\documentclass[12pt]{article}
\usepackage{graphicx}
\usepackage{lineno}
\pdfoutput=1
\newcommand\pubnumber{SNSN-323-63}
\newcommand\pubdate{\today}

\def\institute{Centre de Physique des Particules de Marseille,\\
Aix Marseille Univ, CNRS/IN2P3, CPPM, Marseille, France,\\
163 Av. de Luminy, Case 902, 13288 Marseille Cedex 9, France}
\def\support{\footnote{Copyright 2018 CERN for the benefit of the ATLAS Collaboration. Reproduction of this article or parts of it is allowed as specified in the CC-BY-4.0 license.}}

\def\Title#1{\begin{center} {\Large #1 } \end{center}}
\def\Author#1{\begin{center}{ \sc #1} \end{center}}
\def\Address#1{\begin{center}{ \it #1} \end{center}}

\newcommand\pubblock{\rightline{\begin{tabular}{l} \pubnumber\\
         \pubdate  \end{tabular}}}
\newenvironment{Abstract}{\begin{quotation}  }{\end{quotation}}
\newenvironment{Presented}{\begin{quotation} \begin{center} 
             PRESENTED AT\end{center}\bigskip 
      \begin{center}\begin{large}}{\end{large}\end{center} \end{quotation}}

\def\beq{\begin{equation}}
\def\eeq#1{\label{#1}\end{equation}}
\def\eeqn{\end{equation}}

\def\beqa{\begin{eqnarray}}
\def\eeqa#1{\label{#1}\end{eqnarray}}
\def\eeqan{\end{eqnarray}}

\let\bar=\overbar

\def\Dslash{\not{\hbox{\kern-4pt $D$}}}
\def\dslash{\not{\hbox{\kern-2pt $\del$}}}

\def\msb{{\bar{\ssstyle M \kern -1pt S}}}

\RequirePackage{xspace}
\newcommand{\ttbar}{\ensuremath{t\bar t}}
\newcommand{\ttH}{\ensuremath{t\bar t H}\xspace}

\newcommand{\ttV}{\ensuremath{t\bar t V}\xspace}
\newcommand{\ttW}{\ensuremath{t\bar t W}\xspace}
\newcommand{\ttZ}{\ensuremath{t\bar t Z}\xspace}

\newcommand*{\ifb}{\mbox{fb$^{-1}$}}

\def\SevenTeV{\ensuremath{\sqrt{s}=7~\mathrm{TeV}}}
\def\EightTeV{\ensuremath{\sqrt{s}=8~\mathrm{TeV}}}
\def\ThirTeV{\ensuremath{\sqrt{s}=13~\mathrm{TeV}}}

\newcommand\Htobb{\ensuremath{H\to b\bar{b}}\xspace}
\newcommand\Htott{\ensuremath{H\to \tau \tau}\xspace}
\newcommand\Htoyy{\ensuremath{H\to \gamma \gamma}\xspace}
\newcommand\HtoZZ{\ensuremath{H \to ZZ^{*}}\xspace}
\newcommand\HtoWW{\ensuremath{H \to WW^{*}}\xspace}
\newcommand\Htomulti{$H \to (WW^{*},\tau\tau,ZZ^{*}) \to$ leptons\xspace}
\newcommand\Htofourl{$H \to ZZ^{*} \to 4l$\xspace}

\begin{document}
\begin{titlepage}
\pubblock

\vfill
\Title{Observation of \ttH production with ATLAS}
\vfill
\Author{ Fabrice Hubaut, \\
On behalf of the ATLAS Collaboration\support}
\Address{\institute}
\vfill
\begin{Abstract}
An overview of the analyses that led to the observation of the Higgs boson production in association with a top quark pair (\ttH) with the ATLAS experiment is given. 
Up to 80~\ifb\ of proton--proton collision data recorded by the ATLAS experiment at a center-of-mass energy of \ThirTeV\ at the Large Hadron Collider (LHC) have been used to 
achieve this major milestone.
\end{Abstract}
\vfill
\begin{Presented}
$11^\mathrm{th}$ International Workshop on Top Quark Physics\\
Bad Neuenahr, Germany, September 16--21, 2018
\end{Presented}
\vfill
\end{titlepage}
\def\thefootnote{\fnsymbol{footnote}}
\setcounter{footnote}{0}

\section{Introduction}

All measurements of Higgs boson properties are consistent with the Standard Model (SM) predictions, but the fermion sector was so far less tested. 
In the SM, couplings between the Higgs boson and fermions occur through Yukawa interactions, whose strength is proportional to the fermion mass. 
The top quark Yukawa coupling is therefore predicted to be by far the largest one, with a value close to unity. Indirect constrains on this coupling can be obtained 
via the Higgs boson production by gluon--gluon fusion or its decay to two photons, which proceed through loops. However, New Physics (NP) could also enter in these loops, 
hence a direct measurement is needed to disentangle and probe NP. Such a direct measurement can be achieved through the associated production of the Higgs boson and a top 
quark pair (\ttH), which is a tree-level process and whose cross-section is proportional to the square of the top quark Yukawa coupling. Observing the \ttH production would therefore 
allow to simultaneously constrain this coupling and possible Beyond the Standard Model (BSM) effects.\\

The \ttH production is quite a rare process, with a cross-section at the Large Hadron Collider (LHC) of $\sim 0.5$~pb at a center-of-mass energy of \ThirTeV. 
This is typically 1\% of the inclusive Higgs boson production cross-section and less than 0.1\% of the \ttbar\ production one. This rare production has to be convolved with all possible combinations 
of \ttbar\ and Higgs boson decays, giving rise to a very rich spectrum of possible signatures. A broad spectrum of analyses has therefore been designed to cover these multiple final states,
with their selections designed to avoid any overlap between them. 
They will be described in the next sections together with their combination, starting from the most abundant ones (\ttH, \Htobb and \ttH, \Htomulti), which use 36.1~\ifb\ 
of proton--proton collision data recorded by the ATLAS experiment~\cite{atlas} at a center-of-mass energy of \ThirTeV, to the more rare ones (\ttH, \Htoyy and \ttH, \Htofourl ($l=e, \mu$)), 
which use 79.8~\ifb\ of data.

\section{Measurement in the \Htobb channel}

The search for the \ttH production in the \Htobb channel~\cite{bb} benefits from a large branching ratio, 
but suffers from a huge combinatoric background at the event reconstruction level and a large irreducible background
(mostly \ttbar\ $+$ heavy flavour jets) associated with big theoretical uncertainties. The modelling of this background and the definition of the 
associated uncertainty scheme are performed using state-of-the-art Monte Carlo (MC) generators and constraints from data in background-enriched regions.
The nominal \ttbar\ MC sample uses an inclusive next-to-leading order (NLO) generator normalised to predictions at next-to-next-to-leading order in perturbative QCD including resummation of 
next-to-next-to-leading logarithmic soft gluon terms. The main associated
systematic uncertainties are based on comparisons against alternative MC samples, using different generators, parton shower or hadronisation models.
The normalisations of \ttbar+$\ge1b$ and \ttbar+$\ge1c$ backgrounds are derived from data.
To define the analysis regions, a selection of semi- and di-leptonic \ttbar\ decays is first performed, based on the presence of one or two light leptons, 
followed by a further categorisation based on the jet multiplicity and on the value of the $b$-tagging discriminant of each jet (four $b$-tagging working points are used).
This allows to define 9 signal-enriched regions (SR) and 10 background-enriched control regions (CR), with very different signal purities and background fractions
(a ``boosted'' category targeting high $p_\mathrm{T}$ top/Higgs candidates is also defined). The best signal purity in these regions is only $\sim$5\%, so a cascade
of Multivariate Analysis (MVA) classifiers is used to further separate the signal from the backgrounds. A \ttH system reconstruction Boosted Decision Tree (BDT)
is designed in most SR to solve the combinatorics to identify the best parton-jet assignments and reconstruct the final state. It is complemented
with a Likelihood discriminant in the single-lepton SR, built from all different jet permutations, and a Matrix Element Method in the single-lepton SR with the best purity.
The outputs of all these classifiers, together with the related Higgs/top candidate properties, event kinematic variables and $b$-tag information, are input into a classification BDT
built in each SR to best separate the signal from the backgrounds. A combined profile likelihood fit is performed across all analysis regions to extract the signal yield.
The significance of the observed (expected) signal is $1.4\sigma$ ($1.6\sigma$). The measured signal strength is in good agreement with the SM prediction within the relative measurement 
precision of 75\%, as seen in Figure~\ref{fig_1} (left). This precision is dominated by the systematic uncertainties, in particular those associated with \ttbar+$\ge1b$ modelling.

\section{Measurement in multilepton final states}

The search for \ttH production in multilepton final states~\cite{ml} targets \HtoWW, \Htott and \HtoZZ decays. The \ttbar\ background is suppressed by requiring two same-sign or at least 3 leptons.
Seven orthogonal channels, categorised by the multiplicity of light leptons and hadronically decaying $\tau$ ($\tau_{\mathrm{had}}$) candidates, are defined.
They cover a wide range of yields and S/B purities, with different background composition in the various regions.
The irreducible background, made of SM processes with prompt leptons, is dominated by \ttW, \ttZ and diboson productions. 
It is estimated from NLO MC predictions and validated in data in various dedicated regions.
The reducible background is mainly coming from \ttbar\ events, with non-prompt light leptons from $b$-hadron decay, fake $\tau_{\mathrm{had}}$ candidates from
light flavour jet or charge misidentified electrons. These contributions are mostly estimated through data-driven techniques, and validated in various regions.
Dedicated MVA approaches are used in most SR to further enhance the signal to background separation. A combined profile likelihood fit across all SR and CR extracts
the signal with an observed (expected) significance of $4.1\sigma$ ($2.8\sigma$). The measured signal strength is in good agreement with the SM prediction within the relative measurement 
precision of 30\%, as seen in Figure~\ref{fig_1} (right). This precision is fairly shared between statistical and systematic errors, dominated by uncertainties 
on \ttH and \ttV ($V=W,Z$) modelling and those associated to non-prompt lepton estimates (with a large contribution from the limited CR statistics). 
Similar results are obtained with an alternative fit using a free floating \ttV normalisation, albeit with a 15\% degraded sensitivity.

\begin{figure}[htb]
\includegraphics[width=0.52\linewidth]{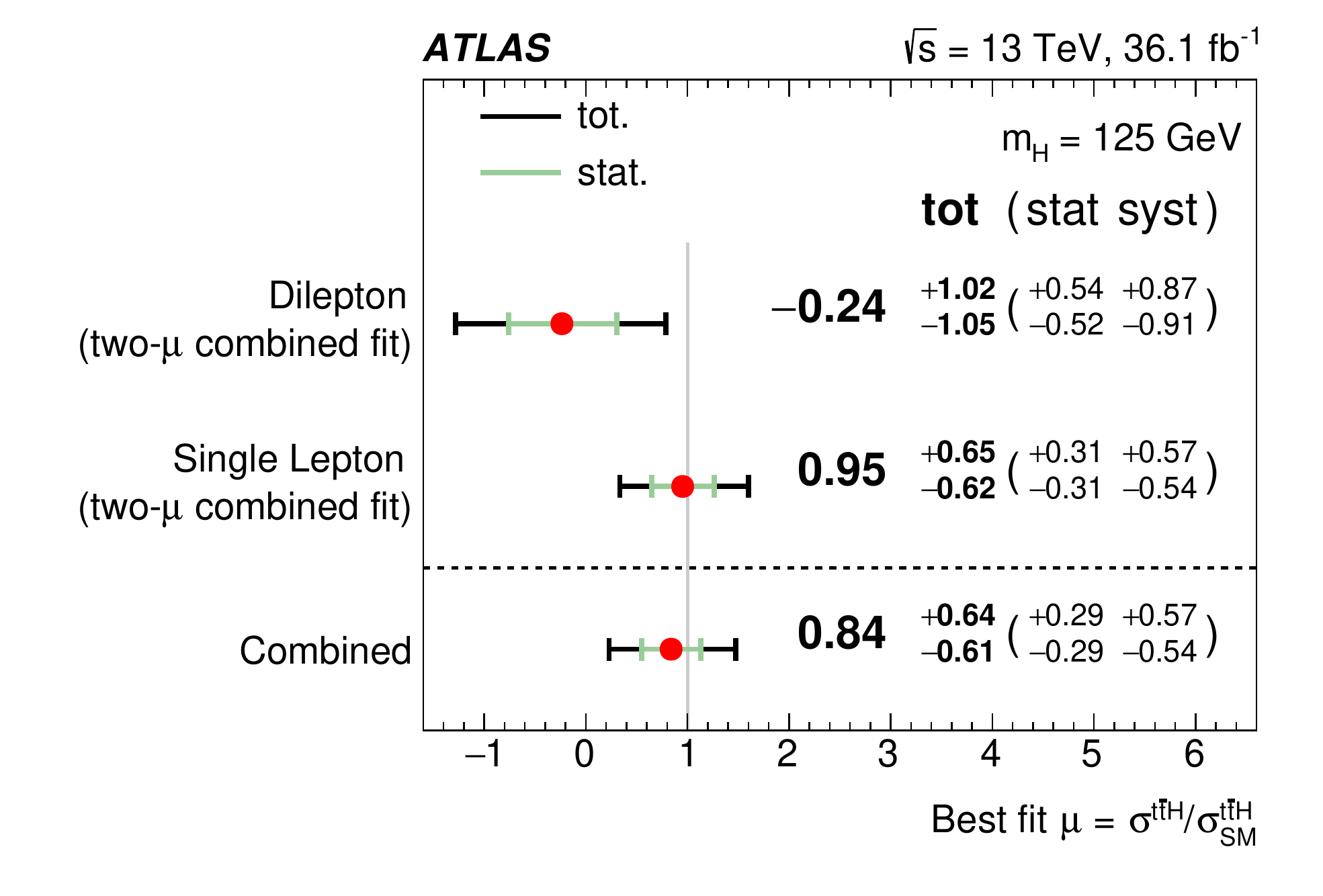}
\includegraphics[width=0.48\linewidth]{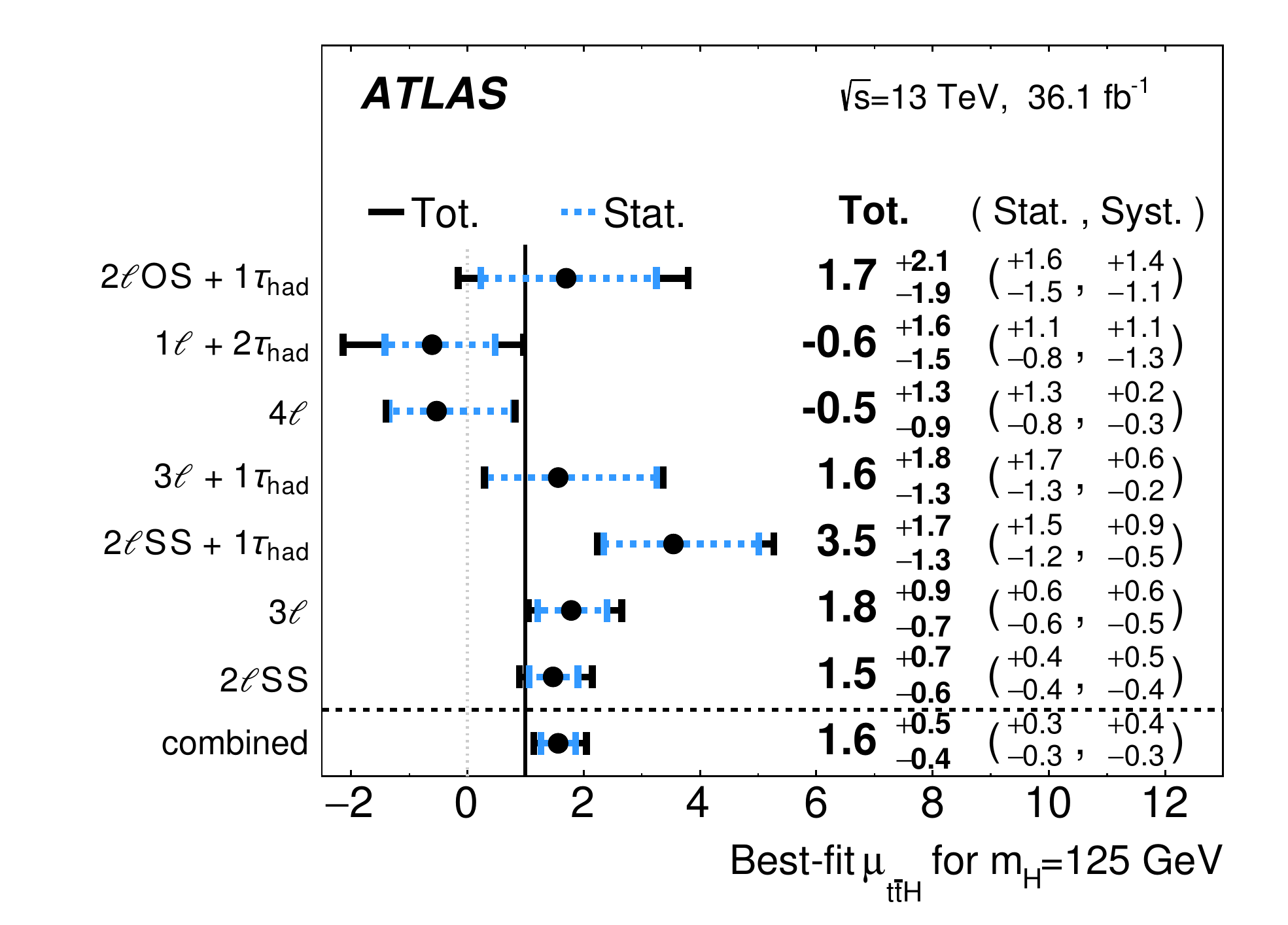}
\vspace*{-1cm}
\caption{Observed best-fit values of the \ttH signal strength $\mu$ and their uncertainties by final-state category and combined, for the \ttH, \Htobb (left~\cite{bb}) and multileptons (right~\cite{ml}) analyses. 
The individual $\mu$ values for the channels are obtained from a simultaneous fit with the signal-strength parameter for each channel floating independently. The SM prediction is $\mu=1$.
\label{fig_1}}
\end{figure}

\section{Measurement in the \Htoyy and \Htofourl ($l=e, \mu$) channels}
The \ttH, \Htoyy channel~\cite{obs} is more rare, but offers a clean signature with a narrow peak in the diphoton invariant mass distribution over a smooth continuum background, 
extrapolated from sideband regions.  Two high-$E_\mathrm{T}$ isolated photons are required, and events are categorised in leptonic 
and hadronic regions depending on the presence of a light lepton. A further signal to background separation power is based on BDT discriminants, 
trained with deep neural networks using \ttH MC and background events from data control regions. Seven categories (4 hadronic and 3 leptonic) are defined according to the value of the BDT output. 
The signal over background ratio is $\sim1.9$ in the purest categories, with the background dominated by the diphoton continuum (including \ttbar$+\gamma\gamma$). 
A combined unbinned fit to the diphoton invariant mass distributions in the 7 categories is performed (see Figure~\ref{fig_2} (left)), with the Higgs signal modelled with a double-sided Crystal Ball function, 
and the continuum background with a functional form whose parameters are free in the fit. The significance of the observed (expected) signal is $4.1\sigma$ ($3.7\sigma$). 
The measured signal strength is in good agreement with the SM prediction within the relative measurement 
precision of 30\%, dominated by statistical errors. The dominant systematic uncertainties are associated to the signal parton shower model (8\%) and the photon energy resolution (6\%).\\

A similar strategy is performed in the \ttH, \Htofourl ($l=e, \mu$) channel~\cite{obs}, with a categorisation in leptonic and hadronic
regions and the use of a BDT discriminant in the hadronic category. The signal over background ratio is $\ge5$ in the purest categories, with the background dominated by \ttV 
and non-\ttH Higgs production modes. The expected significance is $1.2\sigma$ (0.6 \ttH events), while 0 event are observed.

\section{Combination and results}

The combination~\cite{obs} of the searches for \ttH production described in the previous sections is based on a simultaneous fit to all signal and control regions of the individual analyses. 
The contributions from the non-\ttH Higgs production modes are fixed to SM predictions. 
Using \ThirTeV\ data only, an observed (expected) signal significance of $5.8\sigma$ ($4.9\sigma$) is achieved. 
Combining with \SevenTeV\ and \EightTeV\ data increases the observed (expected) signal significance to $6.3\sigma$ ($5.1\sigma$). 
This constitutes an observation of the \ttH process, which was a major milestone for LHC Run-2. 
The total \ttH cross-section at \ThirTeV\ is measured to be $670\pm90$ (stat) $^{+110} _{-100}$ (syst)~fb, in good agreement with the NLO SM prediction of $507^{+35} _{-50}$~fb. 
The measurement precision of 20\% is slightly dominated by systematic uncertainties, in particular from \ttbar+heavy flavour modelling (9.9\%) and signal modelling (6.0\%). 
The cross-section measurements performed at \ThirTeV\ and \EightTeV\ are shown in Figure~\ref{fig_2} (right).\\

Measurements of Higgs boson couplings combining several ATLAS analyses, including those on \ttH reported here, have been performed~\cite{couplings}. 
Assuming SM particles only ($gg\to H$ and \Htoyy loops are then resolved with SM particles), the top quark Yukawa coupling, relative to its SM value, 
is measured to be $\kappa_{top}=1.03^{+0.12} _{-0.11}$. Allowing new particles to enter in the loops, an effective Higgs-to-gluon coupling is defined, 
and a ratio $\kappa_{gluon}/\kappa_{top}=1.09\pm0.14$ is extracted. This is consistent with SM Higgs boson coupling predictions, and allows to put constrains on BSM contributions.

\begin{figure}[htb]
\includegraphics[width=0.55\linewidth]{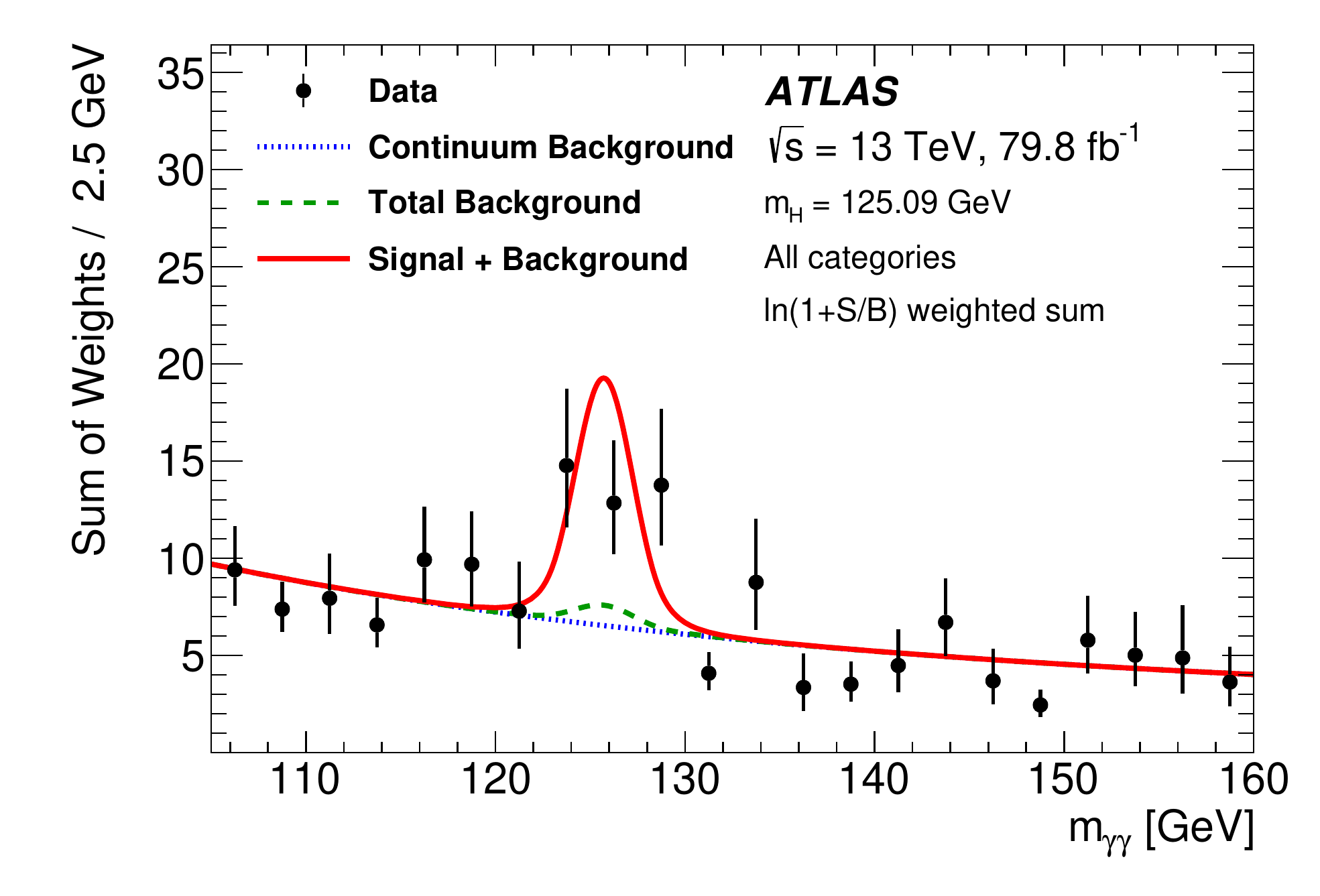}
\includegraphics[width=0.45\linewidth]{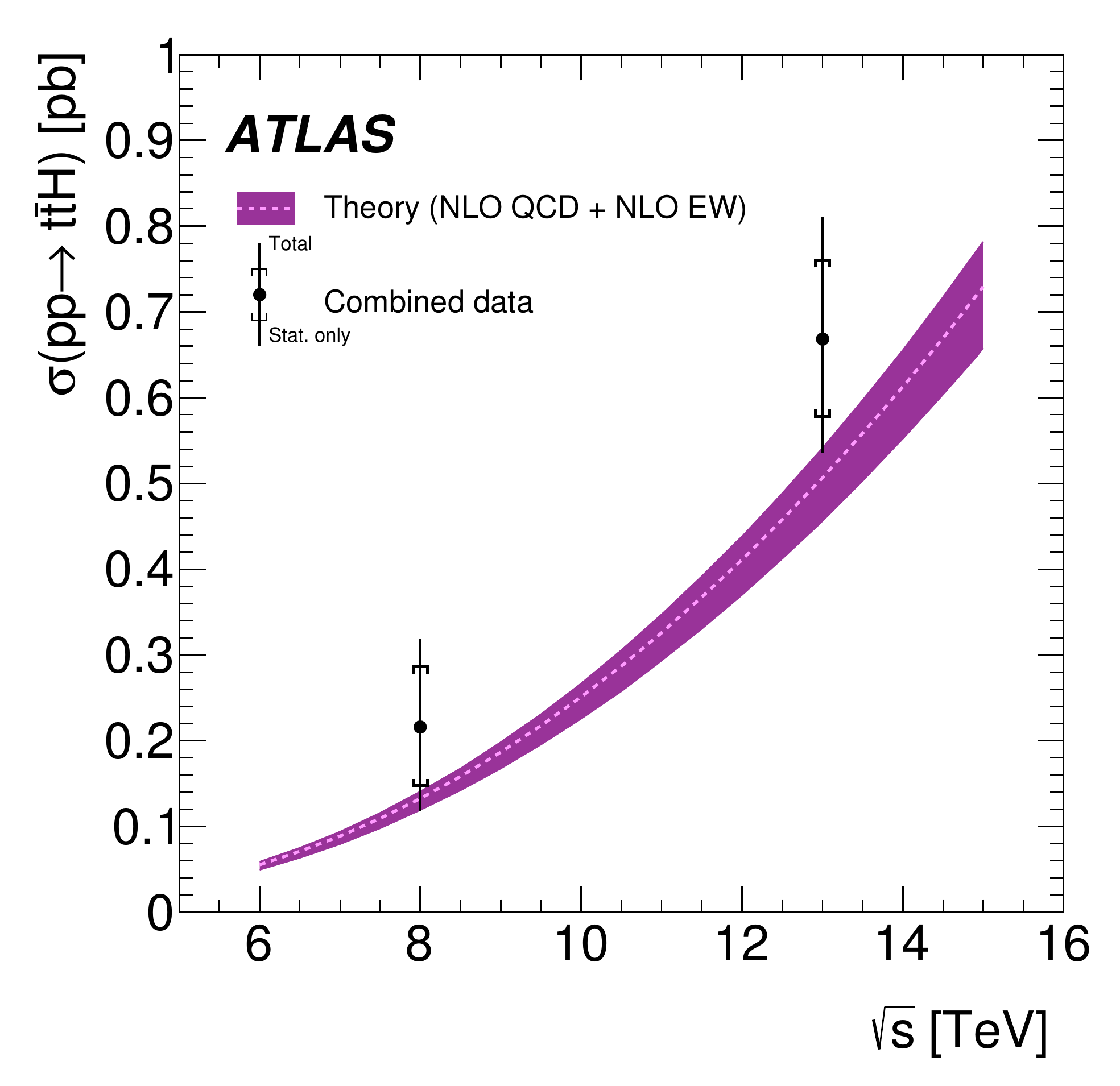}
\vspace*{-1cm}
\caption{Left: Weighted diphoton invariant mass spectrum in the \ttH-sensitive BDT bins. 
The solid red curve shows the fitted signal-plus-background model. 
The non-resonant and total background components of the fit are shown with the dotted blue curve and dashed green curve. 
Right: Measured \ttH cross-sections in $pp$ collisions at center-of-mass energies of 8 TeV and 13 TeV. 
Both the total and statistical-only uncertainties are shown. The measurements are compared with the SM prediction. Ref.~\cite{obs}.
\label{fig_2}}
\end{figure}

\section{Conclusions}

The \ttH production process has been observed using $pp$ collision data produced by the LHC and recorded with the ATLAS detector.
The cross-section at \ThirTeV\ is measured with 20\% precision.
All measurements are in good agreement with SM predictions, allowing to put constrains on BSM contributions.
This constitutes a direct observation of the top quark Yukawa coupling, which, together with the recent observations of \Htott and \Htobb decays
by the ATLAS and CMS collaborations separately, firmly establishes the existence of Yukawa interactions.

\end{document}